\documentclass{aa}
\usepackage{graphicx}
\usepackage{natbib}
\usepackage{amsmath}
\usepackage{amsfonts}
\bibpunct{(}{)}{;}{a}{}{,}

\newcommand{\PFA}{P_{\rm FA}}
\newcommand{\PD}{P_{\rm D}}

\newcommand{\Cb}{\boldsymbol{C}}
\newcommand{\nb}{\boldsymbol{n}}
\newcommand{\ssb}{\boldsymbol{s}}
\newcommand{\tdelta}{\widetilde{\delta}}
\newcommand{\tnu}{\widetilde{\nu}}
\newcommand{\tkappa}{\widetilde{\kappa}}
\newcommand{\xb}{\boldsymbol{x}}

\newcommand{\lhbs}{{\it LHBS~}}

\begin{document}

\title{Some comments on the paper \\ ``{\it {\bf Filter design for the detection of compact sources
based on the Neyman-Pearson detector}}'' \\ by M. L\'opez-Caniego et. al (2005, MNRAS 359, 993)}

   \author{R. Vio\inst{1}
          \and
          P. Andreani\inst{2}
          }

   \offprints{R. Vio}

   \institute{Chip Computers Consulting s.r.l., Viale Don L.~Sturzo 82,
              S.Liberale di Marcon, 30020 Venice, Italy\\
              \email{robertovio@tin.it}
         \and
             INAF-Osservatorio Astronomico di Trieste, via Tiepolo 11,
                  34131 Trieste, Italy \\
             \email{andreani@ts.astro.it}
             }

\date{Received .............; accepted ................}

\abstract{In this note we stress the necessity of a careful check of both the theoretical arguments and 
the numerical experiments used by \citet{lop05} to support the superior performances of the {\it biparametric 
scale adaptive filter} (BSAF) with respect to the classic {\it matched filter} (MF) in the detection of 
sources on a random Gaussian background.
}
\titlerunning{Optimal Detection of Sources}
\authorrunning{R. Vio, P. Andreani}
\maketitle

\section{Introduction}

A popular tool for the detection of sources embedded in a noise background is the 
{\it matched filter} (MF). This is because its {\it optimal} properties. In particular, in the case of 
Gaussian noise, it can be shown that through MF it is possible to construct {\it uniformly most powerful tests}, 
i.e., tests that for a given {\it probability of false alarm} $\PFA$ (the probability of a 
false detection) provide the best {\it probability of detection} $\PD$ (the probability of a 
correct detection). This is guaranteed by the Neyman-Pearson (NM) theorem. For the
sake of clarity and completeness, some arguments are presented in the following.

Let's suppose that $\xb = [ x(0), x(1), \ldots, x(N-1)]^T$ is an observed one-dimensional discrete signal.
In its simplest form, the detection problem consists in deciding whether $\xb$ is constituted by a pure
zero-mean Gaussian noise $\nb$ with covariance matrix $\Cb$  (hypothesis $H_0$)
or it contains also the contribution 
of a deterministic component $\ssb$ (hypothesis $H_1$). Both $\Cb$ and $\ssb$ are supposed known.
In other words, the source detection problem is equivalent to a decision problem where
one has to distinguish between the two hypotheses:
\begin{equation}
\left\{
\begin{array}{ll}
H_0: & \quad \xb = \nb; \\
H_1: & \quad \xb = \nb + \ssb.
\end{array}
\right.
\end{equation}
Under $H_0$ the probability of $\xb$ is given by
\begin{align}
p(\xb| H_0)  &  = p_G(\nb) = p_G(\xb) \nonumber \\
& \equiv \frac{1}{(2 \pi)^{N/2} {\rm det}^{1/2}(\Cb)} \exp\left[-\frac{1}{2} \xb^T\Cb^{-1}\xb\right],
\end{align}
whereas under $H_1$ it is
\begin{align} \label{eq:h1}
& p(\xb| H_1) = p_G(\nb) = p_G(\xb - \ssb) \nonumber \\
& \equiv \frac{1}{(2 \pi)^{N/2} {\rm det}^{1/2}(\Cb)} \exp\left[-\frac{1}{2} (\xb-\ssb)^T\Cb^{-1}(\xb-\ssb)\right].
\end{align}
The choice between $H_0$ and $H_1$ requires the definition of  the so called {\it critical} or {\it acceptance region}
$R_*$. This is set of values in 
$\mathbb{R}^N$ that map into the decision $H_1$ or
\begin{equation}
R_* = \{\xb: \textrm{decide } H_1 \textrm{or reject } H_0 \}.
\end{equation}
Once $R_*$ has been fixed, the quantities $\PD$ and $\PFA$ are given by
\begin{equation} \label{eq:p1}
\PD = \int_{R_*} p(\xb| H_1) d\xb,
\end{equation}
and
\begin{equation} \label{eq:p2}
\PFA = \int_{R_*} p(\xb| H_0) d\xb.
\end{equation}
The Neyman-Pearson (NP) theorem tells us how to choose $R_*$ in such a way to maximize $\PD$
when we are given the likelihood functions $p(\xb| H_0)$ and $p(\xb|H_1)$, and a value $\alpha$ for $\PFA$.

\medskip
\noindent
{\bf Neyman-Pearson theorem:} {\it to maximize $\PD$ for a given $\PFA=\alpha$, decide $H_1$ 
if the {\it likelihood ratio}
\begin{equation} \label{eq:np}
L(\xb) = \frac{p(\xb| H_1)}{p(\xb| H_0)} > \gamma,
\end{equation}
where the threshold $\gamma$ is found from
\begin{equation}
\PFA = \int_{\{\xb: L(\xb) > \gamma\}} p(\xb| H_0) d\xb = \alpha.
\end{equation}
}

\noindent
By inserting Eqs.\eqref{eq:p1}-\eqref{eq:p2} in Eq.\eqref{eq:np}, it is possible to see that
$H_1$ has to be chosen when for the statistic $T(\xb)$ (called {\it NP detector}) it is
\begin{equation} \label{eq:T}
T(\xb) = \xb^T \Cb^{-1} \ssb > \gamma',
\end{equation}
with $\gamma'$ such as
\begin{equation} \label{eq:pfa}
\PFA = Q \left(\frac{\gamma'}{\left[ \ssb^T \Cb^{-1} \ssb \right]^{1/2}} \right) = \alpha.
\end{equation} 
Here, $Q(x) = 1 - \Phi(x)$ with $\Phi(x)$ the {\it cumulative distribution function} of the
standard Gaussian distribution. 
Equation~\eqref{eq:pfa} is due to the fact that, being given by a linear combination of random Gaussian quantities,
$T(\xb)$ is a random Gaussian quantity too. For the same reason it happens that
\begin{equation}
\PD = Q \left( Q^{-1} \left( \PFA \right) - \sqrt{ \frac{\epsilon}{\sigma^2}} \right),
\end{equation}
with $\epsilon = \ssb^T \ssb$ the {\it signal energy} and $\sigma$ the standard deviation of the noise
$\nb$. In Eq.\eqref{eq:T}, the quantity $\Cb^{-1} \ssb$ provides the classical MF. Therefore, this filter
can be expected to have {\it optimal} properties. An important point to stress is that these properties
do not degrade very much when the amplitude and/or the position 
of the source are unknown and have to be estimated from the data themselves \citep{kay98}.

In \citet{lop05} (hereafter \lhbs) an alternative approach is proposed that is claimed to provide filters
with superior performances with respect to MF. Authors work in the context of one-dimensional continuous signal and under 
the hypothesis of sources with symmetric profile. Their idea consists in exploiting the joint distribution 
$p(\nu_n, \kappa_n)$ of the normalized maxima $\nu_n$ and of the corresponding
normalized curvature $\kappa_n$ that, for a continuous Gaussian random noise, is given by the Rice's formula
\begin{equation} \label{eq:PH0}
p_R(\nu_n, \kappa_n) = \frac{\kappa_n}{\sqrt{2 \pi (1 -\rho^2)}} \exp\left[ - \frac{\nu^2_n + \kappa_n^2 - 2 \rho 
\nu_n \kappa_n}
{2(1-\rho^2)} \right].
\end{equation}
Here, $\nu_n \equiv \xi / \sigma_0 \in (-\infty, +\infty)$, $\kappa_n \equiv - \xi'' / \sigma_2 \in (0, +\infty)$ 
with $\xi$ the value of the maxima
of the signal and $\xi''$ the corresponding second derivative, whereas $\sigma_i^2$ and $P(q)$ are the moment of order
$2 i$ 
\begin{equation}
\sigma^2_i \equiv 2 \int_0^{\infty} q^{2i} P(q) dq,
\end{equation}
and the power spectrum of the noise process, respectively. Of course, for the first normalized derivative
of the maxima it is $\delta_n = \xi'/\sigma_1 = 0$.
 
If a source is embedded in the background and $\nu$, $\delta$ and $\kappa$ are observed quantities,
the NP approach provides the following decision problem
\begin{equation} \label{eq:decision2}
\left\{
\begin{array}{ll}
H_0: & \quad \nu = \nu_n, \quad \delta = \delta_n = 0, \quad \kappa = \kappa_n; \\
H_1: & \quad \nu = \tnu_n + \tnu_s, \quad \delta = \tdelta_n + \tdelta_s=0, \quad \kappa = \tkappa_n + \tkappa_s,
\end{array}
\right.
\end{equation}
where $\tnu_n$, $\tnu_s$ are the values of the noise process and of the source in correspondence
to the position of the observed maximum, whereas $\tdelta_n$, $\tdelta_s$ and $\tkappa_n$, $\tkappa_s$ are the corresponding normalized first derivatives and curvatures, respectively. In other words, 
under $H_0$ it is $p(\nu, 0, \kappa| H_0) = p_R(\nu_n, \kappa_n)$, whereas under $H_1$
it is $p(\nu, 0, \kappa| H_1) = p_f(\tnu_n, \tdelta_n, \tkappa_n)$.
A point to stress is that
the likelihood $p_f(\tnu_n, \tdelta_n, \tkappa_n)$ is not given by the Rice's formula. 
This is because, in general,  there is no guarantee that the position of the observed maximum coincides
with the position of the peak of the source and/or with the position of one of the peaks in the noise process.

\section{Some comments on the \lhbs approach}

The \lhbs approach deserves some comments. Among these, three are of particular relevance. The first one is that,
instead of $p_f(\tnu_n, \tdelta_n, \tkappa_n)$, under $H_1$ authors use a likelihood in the form
\begin{align} \label{eq:PH1a}
& p(\nu, \kappa| H_1) = \frac{\kappa}{\sqrt{2 \pi (1 -\rho^2)}} \nonumber \\
& \times \exp\left[ - \frac{(\nu^2 - \nu_s)^2 + 
(\kappa-\kappa_s)^2 - 2 \rho 
(\nu-\nu_s) (\kappa-\kappa_s)}
{2(1-\rho^2)} \right],
\end{align}
where $\nu \in (-\infty, +\infty)$ and $\kappa \in (0, +\infty)$. Actually, 
it is not clear from where this equation comes out. In fact, even in the unrealistic hypothesis that the
position of the peak of the source has to coincide with the position of a maximum of the noise process, similarly
to Eq.\eqref{eq:h1} it should be
\begin{align} \label{eq:PH1b}
& p(\nu, \kappa| H_1) = \frac{\kappa - \kappa_s}{\sqrt{2 \pi (1 -\rho^2)}} \nonumber \\
& \times \exp\left[ - \frac{(\nu^2 - \nu_s)^2 + 
(\kappa-\kappa_s)^2 - 2 \rho 
(\nu-\nu_s) (\kappa-\kappa_s)}
{2(1-\rho^2)} \right],
\end{align}
with $\nu \in (-\infty, +\infty)$ and $\kappa \in (\kappa_s, +\infty)$. 
\lhbs justify the use of the likelihood~\eqref{eq:PH1a} only with a reference
to \citet{bar03}. However, also in that work no details are provided.

The second comment regards the fact that \lhbs do not
work on the original signal $\xb$, but on a filtered version of it. The reason is that
this operation modifies $R_*$ \citep[see Fig.5 in ][]{bar03}. 
Hence, \lhbs try to define a filter for which, once 
fixed $\PFA$, $R_*$ is such as to maximize the corresponding $\PD$. This is accomplished by parametrizing 
some well known filters (e.g., MH and the {\it Mexican hat wavelet}) and by computing the values
of the parameters through an optimization procedure. For example, the {\it biparametric scale adaptive filter} 
(BSAF), claimed by  \lhbs to outperform MH, is a modified version of MF where two coefficients are assumed to be
free parameters. However, authors provide only an heuristic
explanation of the reason why such an approach should be expected to work. No rigorous argument
is given. Probably, the conjecture is that their procedure provides a two-steps 
(parameter optimization + filtering) approximation 
of the solution of the decision problem based on the {\it likelihood ratio}
\begin{equation} \label{eq:lr}
L(\xb,\nu, \kappa) = \frac{p(\xb, \nu, \kappa| H_1)}{p(\xb, \nu, \kappa| H_0)} > \gamma,
\end{equation}
that (at least in principle) should better exploit the available {\it a priori} information. 
The main problem to work directly with such {\it likelihood ratio} is in the difficulty to obtain the joint
distributions $p(\xb, \nu, \kappa| H_0)$ and  $p(\xb, \nu, \kappa| H_1)$ that can be quite complex.
In any case, the usefulness of the proposed procedure is not
obvious and should be proved through rigorous theoretical arguments. The conclusions drawn only on the basis of 
numerical experiments lack 
generality and can be potentially misleading. This is especially true for the discrete case where the 
quantities $\nu$ and $\kappa$ have to be computed by using the data $\xb$.

The last comment regards the procedure followed by authors in the numerical experiments. In particular,
it is questionable that they investigate the performance of their detection test only on the
sources that, after the filtering, show a peak placed at the position of their true maximum. 
This is certainly an unrealistic condition. Moreover, in this way there is
the risk to favour the selection of the sources that are located in correspondence to a maximum of 
the noise, i.e., those whose likelihood $p(\nu, \kappa| H_1)$ is given by Eq.\eqref{eq:PH1b}. In the discrete case, 
it easy to realize that this is very probable when, for example, a source is superimposed on a 
white noise background and no filtering operation is carried out. On the other side, BSAF is a 
symmetric filter that works principally on spatial scales smaller than a pixels. Hence, the necessity
to check for the above mentioned risk. The correct procedure requires that the detection test
be applied also to the sources that after the filtering show a maximum at a position different
from the true one.

\section{Conclusions}

From the discussion presented above, it is evident the necessity of a careful check of both the
theoretical arguments and the numerical experiments used by \lhbs before BSAF 
can be claimed to be superior to the classic MH in source detection problems. 

As last comment, it is useful to stress that the BSAF corresponding to a specific problem can be obtained only
through a very involved procedure based on stringent {\it a priori} assumptions (e.g., Gaussianity of the 
background, symmetry of the source profile). All this makes the procedure proposed by \lhbs
quite rigid. In this respect, MF certainly represents a much more flexible solution.

\end{document}